\documentclass[twoside]{dis08}
\usepackage[latin1]{inputenc}
\usepackage[dvips]{graphicx,epsfig,color}
\usepackage{wrapfig,rotating}
\usepackage{amssymb,amsmath,array}

\begin{document}
\title{The e+A programme at a future Electron-Ion Collider facility}

%***********************************************************************
% AUTHORS INFORMATION AREA
%***********************************************************************
\author{Matthew A. C. Lamont$^1$ for the EIC working group
%
% Optional short acknowledgment: remove next line if non-needed
%\thanks{This is an optional funding source acknowledgment.}
%
% DO NOT MODIFY THE FOLLOWING '\vspace' ARGUMENT
\vspace{.3cm}\\
%
% Addresses and institutions (remove "1- " in case of a single institution)
1- Physics Department - Brookhaven National Laboratory \\
Upton, NY 11973, USA
%
% Remove the next three lines in case of a single institution
}
%***********************************************************************
% END OF AUTHORS INFORMATION AREA
%***********************************************************************

\maketitle

\begin{abstract}

At small $x$, the gluon distribution dominates the nuclear wave function.  The increase needs to be tamed in avoid violating unitarity constraints.  The most efficient way to study this in colliders is through $e$+A collisions as the nucleus is an efficient amplifier of the physics of high gluon densities.   To this end, there are proposals to build an $e$+A machine in the USA which would operate over a large range of energies and masses.  These studies would also allow an in-depth comparison to A+A collisions where recent results have given tantalising hints of a new state of matter produced with partonic degrees of freedom.  As gluon interactions are the dominant source of hard probes, they themselves must be understood before the results are explained quantitatively.  

\end{abstract}

\section{Understanding the Gluon Distributions in Nuclei}

\begin{wrapfigure}{r}{0.5\columnwidth}
\centerline{\includegraphics[width=0.45\columnwidth]{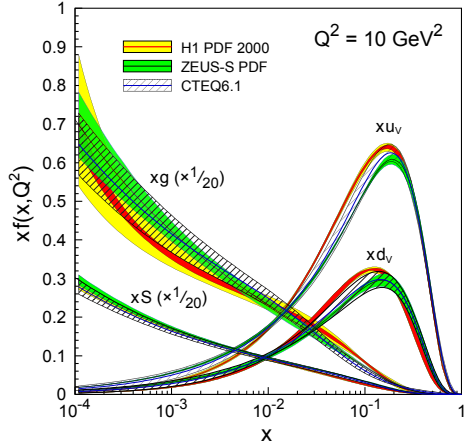}}
\caption{Gluon, sea- and valence-quark momentum distributions in the nucleon taken from a NLO DGLAP fit to $F_2$ measured at HERA (taken from \cite{Saxon:2007zz}).}
\label{Fig:GluonSat}
\end{wrapfigure}

Although all of the unique features of QCD are determined by the self-interactions of gluons, currently, very little is known about their space- and momentum-distributions in nuclei.  As the gluonic degrees of freedom are missing in the hadronic spectrum, in order to study the gluon structure of the nuclear wave-function, high-energy probes of the nucleus are required.  Whilst $p$+A collisions provide excellent information on the gluon properties, as many observables require gluons to participate at the leading order, interpreting the data is difficult due to the soft colour interactions between the $p$ and the A before the hard scattering takes place.

Therefore, the most desirable collision system to probe the gluon properties are lepton+A collisions.  The lepton beams interact with the electrically charged quarks in a process known as Deep-Inelastic Scattering (DIS) and the gluonic part of the nuclear wave-function modifies the interaction in ways which allow the extraction of the gluon properties.  The invariant cross-section in DIS can be written as:

\begin{eqnarray*}
\frac{d^2 \sigma^{eA \rightarrow eX}}{dx dQ^2} =
\frac{4 \pi \alpha^2_{e.m.}}{xQ^4} \left[ \left(1-y+\frac{y^2}{2} \right )
    F^A_2(x,Q^2) - \frac{y^2}{2} F^A_L(x,Q^2) \right]
\end{eqnarray*}

where $y$ is the fraction of the energy lost by the lepton in the rest frame of the nuclei.  $F^A_2$ represents the quark and anti-quark structure function and $F^A_L$ represents that of the gluons.  $F_2$ was studied extensively at HERA for protons, where the gluon properties were inferred through the scaling violation of this structure function.  A direct measurement of $F_L$ is more complicated and requires data at different energies.  This was achieved at HERA in the 2007 run and first results were reported by both H1 and ZEUS at this conference~\cite{Ref:HERA}.

\subsection{Gluon Saturation}

A well known phenomena emerging from DIS experiments on protons at HERA have shown that for $Q^2 \gg \Lambda^2$, the gluon density in the nucleon increases rapidly with small-$x$ and is substantially greater than that of both the valence and the sea quarks for $x <$ 0.01.  This is shown in Fig. \ref{Fig:GluonSat}.  Note that the sea-quark and gluon distributions have been scaled down by a factor of 20.  DIS experiments on nuclei have shown that the quark and gluon distributions are modified compared to their distributions in nuclei at large $x$, known as shadowing, whereas for smaller $x$ ($<$ 0.01), there are no existing measurements of the gluons.  At large $x$ and $Q^2$, the gluon properties are determined by linear evolution equations (DGLAP~\cite{Gribov:1972ri} along $Q^2$ and BFKL~\cite{Kuraev:1977fs} along $x$).  The rapid increase in gluon densities at small $x$ is believed to arise from gluon Bremmstrahlung - hard gluons shedding successively softer gluons.  At small values of $x$, gluon saturation occurs, where this process is matched by the recombination of the excess of soft gluons into harder gluons.  The high number of gluons means that there dynamics are classical and their piling up at the $Q_s^A$ momentum scale is reminiscent of a Bose-Einstein condensate, leading to suggestions that the matter in nuclear wave functions at high energies is universal and can be described as a Colour Glass Condensate~\cite{Ref:CGC}.

%\subsubsection{The Nuclear ``Oomph Factor"}

\begin{wrapfigure}{r}{0.5\columnwidth}
\centerline{\includegraphics[width=0.45\columnwidth]{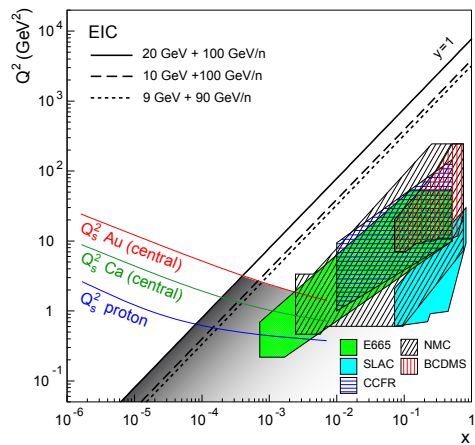}}
\caption{The kinematic acceptance in $x$ and $Q^2$ for different energies realisable at an EIC.  The $Q_s^2$ scale is shown for different species.}
\label{Fig:SatScale}
\end{wrapfigure}

This process is described by the non-linear, small-$x$ renormalization group equations, JIMWLK~\cite{Ref:JIMWLK}.  The onset of this saturation is described by a dynamical scale, $Q_s^2$, which grows with smaller $x$ (larger energy) and increasing nuclear size, meaning that it is experimentally more accessible in heavy  $e$+A collisions than in $e$+$p$ collisions.  Simple estimations suggest that the saturation scale, $Q_s^2 \propto (A/x)^{1/3}$, meaning that the nucleus works as an efficient amplifier of the physics of high gluon densities.  Recent calculations support this argument, even increasing the dependence on A by a further 10$\%$~\cite{Ref:Oomph}.  Therefore, nuclear DIS for heavy nuclei probes the same physics as DIS on protons  as values of $x$ two orders of magnitude lower (or an order of magnitude higher in energy).  For large nuclei, there is a significant window at low x where $Q_s^2 \gg Q^2 \gg\Lambda_{\rm QCD}^2$ and one is in the domain of strong non-linear fields.  This region has not yet been accessible in previous $l$+A collisions, as shown in Fig. \ref{Fig:SatScale}.

\subsection{Experimental Observables} 

The following key questions define the experimental observables in e+A physics:

\textbf{What are the momentum distributions of gluons and sea-quarks in nuclei?}  This measurement is one of the first key measurements at an EIC and $G^A(x, Q^2)$ can be extracted a number of ways: $i$) through the scaling violation of the the quark structure functions, $F_2^A$, with $Q^2$ ($\partial F_2^A/\partial \ln(Q^2) \neq 0$) in the way which was used at HERA for protons, $ii$) through direct measurement for $F_L^A$, obtainable through running at more than one energy, $iii$) through the measurement of inelastic and $iv$) diffractive vector meson production.

\textbf{What are the space-time distributions of gluons and sea-quarks in nuclei?} Not only do we want to understand the momentum distribution of the glue, we also wish to understand the spatial distribution (gluon density profile) in order to understand the physics of high parton densities.  In order to learn about this, we shall use HERA techniques for measuring vector meson production as these are directly applicable to $e$+A collisions.

\textbf{What is the role of Pomerons (colour neutral excitations) in scattering off nuclei?}  The role of diffractive physics, where the electron probe interacts with a Pomeron, is an important interaction in $e$+A collisions, comprising up to 30-40$\%$ of the total cross-section.  Studies of coherent diffractive scattering are easier in a collider environment and performing these measurements at an EIC will allow us to probe directly the structure of the Pomeron and will provide stringent tests on strong gluon field dynamics in QCD. 

\textbf{How do fast probes interact with an extended gluonic medium?}  In DIS on light nuclei, a suppression of hadron production has been observed which is analogous to, but smaller than, that observed at RHIC.  Using nuclear DIS, one can study in detail the energy loss of particles traversing though ``cold nuclear matter".  Experimental data shows that the ratio of hadrons per nuclear DIS event as a function of virtual photon energy, compared to that in deuterium is significantly reduced~\cite{Ref:HERMESEnergyLoss}.  Both energy loss models and pre-hadron absorption models have been applied to the data with limited successes.  At an EIC, we would be able to perform these measurements for a much wider range of virtual photon energies and, crucially, be able to perform these measurements on charmed hadrons.

\section{Connection to Relativistic Heavy Ion Physics}

The strong flow of hadrons has been observed in Au+Au collisions at RHIC, which, for the first time, is in agreement with ideal hydrodynamical models and is much greater than that which hadron-gas models can produce, indicative of a strongly-coupled medium~\cite{Ref:STARWhitePaper}.  These hydrodynamic models suggest that the system produced in RHIC collisions reaches almost complete thermalization by 1 fm/$c$ after the collision.  The mechanisms which lead to this rapid thermalization are currently not understood as there is no information from QCD on thermalization from first principles though it is believed that it is driven by low-x gluons with $k_T^2 < Q_s^2$, where $Q_s$ is the saturation scale to be discussed later.  Understanding this thermalization process will require knowledge of the momentum and spatial distributions of gluons in nuclei, $G_A(x,Q^2,b)$.

Also, the higher cross section at RHIC energies has meant that hard probes have played a crucial role for the first time in heavy-ion collisions.  RHIC data has shown some surprising features.  One of the first hard-probe measurements was the attenuation of high-$p_T$ particles in general and more specifically, the disappearance of back-to-back jets in central collisions, indicating energy loss as they traversed the strongly coupled dense matter.  When looking differentially at particle species, it was found that the charmed hadrons were as suppressed as the light hadrons, a process which was not expected due to the ``dead cone effect"~\cite{Ref:DeadConeEffect}.  This effect is challenging to theory, requiring a re-assessment of the role of collisional energy loss and pre-hadron absorption in cold-nuclear matter.  First results on this effect in $e$+A collisions for light hadrons have been shown by HERMES~\cite{Ref:HERMESEnergyLoss} but a study for heavy flavours and over a wide range of energies still needs to be performed.  

Gluons play an important role in the production of both heavy flavours and jets at RHIC and in the future the LHC.  However, their calculations are based on parton distribution and fragmentation functions obtained from more elementary collisions.  Already collated results show that these distributions are modified in a nuclear medium, with shadowing and saturation playing a role at low $x$ and the EMC effect at higher $x$.  In order to study these effects in fine detail, we require an understanding of gluon distributions in a nuclear medium.

\section{Summary}
Precision measurements of $e$+A collisions at an EIC will open up a new window of study of gluons and their self-interactions - the defining feature of QCD.  The understanding of the gluon momentum and spatial contributions to nuclear structure is an imperative first step in quantitatively understanding the recent results from RHIC.  These measurements can be realised at an EIC, providing high-luminosity high-energy collisions over a wide range of A, providing for a study of QCD beyond the scope of current accelerators.  Please read elsewhere for further details on the EIC accelerator designs~\cite{Ref:Bernd} and more details on the $e$+A and $e$+$p$ programmes~\cite{Ref:RajuWerner}.

\section{Acknowledgements}
I would like to thank my BNL colleagues Thomas Ullrich and Raju Venugopalan for their insightful comments whilst preparing this manuscript.  This work was supported by BNL LDRD grant 07-006.
 
% ****************************************************************************
% BIBLIOGRAPHY AREA
% ****************************************************************************

\begin{footnotesize}
% IF YOU DO NOT USE BIBTEX, USE THE FOLLOWING SAMPLE SCHEME FOR THE REFERENCES
% ----------------------------------------------------------------------------

% ----------------------------------------------------------------------------

\end{footnotesize}

% ****************************************************************************
% END OF BIBLIOGRAPHY AREA
% ****************************************************************************

\end{document}